\documentclass{jpp_mod}

\usepackage[T1]{fontenc}
\usepackage[utf8]{inputenc}
\usepackage{upgreek}
\usepackage{graphicx}
\usepackage{epstopdf, epsfig}
\usepackage{hyperref}
\usepackage{amsmath}
\usepackage{amssymb}
\usepackage{mathabx}
\usepackage{color}
\usepackage{physics}
\usepackage{todonotes}

\newcommand{\erfc}{{\rm erfc}}

\newcommand{\En}{\mathcal{E}}
\renewcommand{\p}{\partial}
\newcommand{\eps}{\epsilon}
\newcommand{\epsn}{\epsilon_0}
\newcommand{\nn}{n_0}
\newcommand{\ii}{\mathrm{i}}
\newcommand{\ee}{\mathrm{e}}
\newcommand{\lD}{\lambda_\mathrm{D}}
\newcommand{\cs}{c_{\mathrm{s}}}
\newcommand{\nus}{\nu_\ast}
\newcommand{\Ma}{\mathcal{M}}
\newcommand{\Fm}{\mathcal{F}}
\newcommand{\phia}{\phi_{\rm A}}
\newcommand{\phimin}{\phi_{\rm min}}
\newcommand{\phimax}{\phi_{\rm max}}
\newcommand{\vth}{v_{\text{th}}}
\newcommand{\gkyl}{\texttt{Gkeyll}}
\newcommand{\CLBO}{C_{\text{LBO}}}

\shorttitle{Collisional age effects on kinetic shocks}
\shortauthor{A.~Sundstr\"{o}m, J.~Juno, J.M.~TenBarge 
  and I.~Pusztai}

\title{Effect of a weak ion collisionality
  on the dynamics of kinetic electrostatic shocks}

\author{Andr\'{e}as~Sundstr\"{o}m\aff{1}
  \corresp{\email{andsunds@chalmers.se}},
  James~Juno\aff{2},
  Jason~M.~TenBarge\aff{3,4} 
 \and Istv\'{a}n Pusztai\aff{1}}

\affiliation{\aff{1}Department of Physics, Chalmers University of
  Technology, 41296 Gothenburg, Sweden
\aff{2}Institute for Research in Electronics and Applied Physics,
University of Maryland, College Park, MD 20742, USA
\aff{3} Department of Astrophysical Sciences, Princeton University,
Princeton, NJ 08543, USA
\aff{4} Princeton Plasma Physics Laboratory, Princeton, NJ 08543, USA}

\begin{document}

\maketitle
\begin{abstract}
In strictly collisionless electrostatic shocks, the ion distribution
function can develop discontinuities along phase-space separatrices,
due to partial reflection of the ion population. In this paper, we
depart from the strictly collisionless regime and present a
semi-analytical model for weakly collisional kinetic shocks. The model
is used to study the effect of small but finite collisionalities on
electrostatic shocks, and they are found to smooth out these
discontinuities into growing boundary layers. More importantly, ions
diffuse into and accumulate in the previously empty regions of phase
space, and, by upsetting the charge balance, lead to growing
downstream oscillations of the electrostatic potential. We find that
the collisional age of the shock is the more relevant measure of the
collisional effects than the collisionality, where the former can
become significant during the lifetime of the shock, even for weak
collisionalities.

\end{abstract}

\section{Introduction}
\label{sec:intro}
Collisionless plasma shock waves are important in numerous
space and astrophysical phenomena \citep{caprioli,Karimabadi14} and in
laboratory experiments \citep{romagnani}. In particular, not only are
they believed to be responsible for cosmic ray acceleration
\citep{bell2013}, but their ability to energize ions makes them a
possible candidate for laser plasma based generation of high-energy
ion beams with a narrow energy spectrum
\citep{HaberbergerNat,tikhonchuk}. In practice, it is highly
non-trivial to produce a quasi-monoenergetic peak in the ion energy
spectrum that competes in total beam charge with the more robust
target normal sheath acceleration mechanism. However, recent
experimental results by \citet{Pak-etal2018} demonstrate that a
shock-accelerated quasi-monoenergetic beam with a high total charge
can be achieved by optimizing plasma profiles.

Often, the Coulomb mean free paths of the plasma particles are much
larger than the width of the shock front or its dynamically relevant
vicinity. In this case, the abrupt change in plasma parameters between
upstream and downstream is set up by some collisionless kinetic
process \citep{Marcowith,tidmankrall}, such as ion reflection in the
electrostatic shock potential \citep{Moiseev63} or electromagnetic
turbulence \citep{bret2013}, and collisions do not play a significant
role in the dynamics of the shock \citep{baloghbook}.  In the other
extreme, when the mean free paths are shorter than the spatial
structures of the shock -- with relevance to inertial confinement
fusion (ICF) and other high-energy-density applications -- the dynamics
is very similar to fluid shocks, where the entropy generation of the
shock is caused by binary collisions. The collisional limit is often
studied by single-fluid hydrodynamic codes, while the collisionless
limit is mostly studied by particle-in-cell (PIC) simulation codes.

The intermediate region of parameters where the physics is kinetic but
collisions also play a role is, however, much less explored than the
above mentioned extremes, although it can be relevant for laser plasma
experiments. Approaching from the high collisionality direction, the
ICF community has already started to explore this region using
Fokker-Planck solvers \citep{thomas2012,keenan2018}. Other types of
laser plasma experiments, such as those aimed at ion acceleration, do
not compress the target but heat it considerably, which corresponds to
small collisionalities. However, when lasers predominantly couple
their energy to the electrons, and the target is of solid density
and/or consists of a high charge number element, the ion
collisionality need not be vanishingly small. Among the limited number
of studies on the effect of a finite collisionality in electrostatic
shocks relevant for ion acceleration experiments, \citet{Turrell2015}
found that in multi-species plasmas the collisional friction between
the different ion species can lead to very rapid heating.

In the field of laser plasmas -- perhaps owing to the short time scales
of the studied phenomena and that the systems considered are
open -- the fact that collisionality need not be order unity to
significantly affect the dynamics of the system is somewhat
overlooked. Meanwhile, there is a wealth of examples in the field of
magnetic confinement fusion where weak but finite collisionality is
essential. The ability to move particles across phase-space
separatrices that divide regions with qualitatively different particle
dynamics -- e.g.\ a trapped--passing boundary, a boundary of a loss cone,
or the threshold of the runaway region -- can make collisions
important, even when they are rare
\citep{dreicer1960,chankin93,nemov99,FPH}.

In this paper, we depart from the strictly collisionless limit, and
investigate the effects of a weak but finite ion collisionality on
kinetic electrostatic shocks.  The process we focus on here, the
collisional population of the originally empty trapped regions of the
ion phase space, is a cumulative effect. Therefore, the relevant
quantity is the \emph{collisional age}, which can reach order-unity
values during the lifetime of the shock, even if the collisionality is
small. We consider laminar (i.e.\ non-turbulent) shocks that exist at
low (i.e.\ order-unity) Mach numbers \citep{sagdeev1966}, and are
relevant for plasma-based ion acceleration experiments
\citep{HaberbergerNat}.

The paper is organized as follows. Section~\ref{sec:model} starts by
describing the assumptions of the model we use to calculate the effect
of collisional scattering of ions across the phase-space separatrices
in kinetic electrostatic shocks, and qualitatively explains the
emerging physical picture. Then, in sections~\ref{sec:pert} and
\ref{sec:diff}, the perturbative orbit-averaged treatment of
collisions and the reduction of the problem to a diffusion equation
are detailed, respectively. Finally, we present the results in
section~\ref{sec:results}, mostly concerning the development of the
potential structure with collisional age for various Mach numbers and
electron-to-ion temperature ratios.

\section{The kinetic shock model}
\label{sec:model}

\begin{figure}
\centering
\input{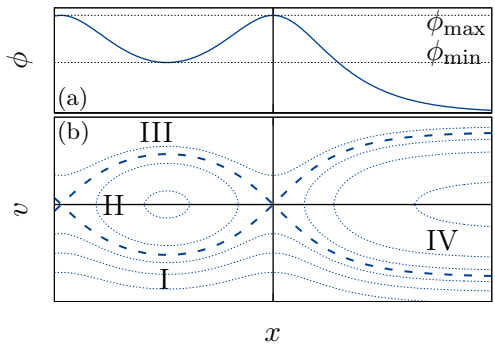}
\put(-178,98){(a)}
\put(-178,85){(b)}
\centering
\caption{(a): Electrostatic potential, $\phi(x)$, of a typical kinetic
  electrostatic shock propagating to the right. The shock has a
  ramp-up in potential, to a maximum of $\phimax$; behind that, the
  potential oscillates between $\phimin$ and $\phimax$.
  (b): Phase space diagram, showing constant energy contours in the
  frame of the shock. The dashed curves denote the upper and lower
  separatrices. Regions of phase space: I -- passing, II -- trapped,
  III -- co-passing, and IV -- reflected.  }
\label{fig:phasespace}
\end{figure}

An important process in the physics of electrostatic shocks is the
partial reflection of ions at the shock front. The electrostatic
potential, see figure~\ref{fig:phasespace}a, ramps up from $\phi=0$ in
the far upstream to a maximum of $\phimax$. The reflection of ions
creates an asymmetry between the up- ($x\ge0$) and downstream ($x<0$)
regions of the shock. This asymmetry, together with the potential
response of the ions and electrons, creates a downstream potential
that oscillates between $\phimax$ and $\phimin$ \citep{sagdeev1966}. 

In this paper, we use the following normalization scheme:
The velocity is normalized to the sound speed $\cs=\sqrt{Z T_{\ee}/m}$
defined with the far upstream electron temperature\footnotemark{},
where the plasma is completely unaffected by the shock, $T_{\ee}$;
$m$ and $Ze$ denote the ion mass and charge, with $e$ the elementary
charge. In particular, the ion flow velocity becomes equal to minus
the Mach number $\Ma=V_0/\cs$, where $V_0$ is the shock speed with
respect to the unperturbed upstream medium. The potential $\phi$ is
normalized to $T_{\ee}/e$, the configuration coordinate $x$ to the
Debye length $\lD=\sqrt{\epsn T_{\ee}/(e^2\nn)}$, and time $t$ to
$\lD/\cs$; here $\nn$ is the electron density far upstream,  where the
plasma is completely unperturbed by the shock, which we normalize to
$1$. All the calculations in this paper are done in the reference
  frame of the shock front.
\footnotetext{The ion temperature is neglected here, since the
  electron-to-ion temperature ratio is assumed to be large, which is
  required for the existence of these types of shocks
  \citep{CairnsPPCF}.} 

The one-dimensional (1D) collisionless shock problem has a steady
state solution, which has been considered in
\citep{pusztaishock}; its solution, in the frame of the shock, is
derived from the time-independent Vlasov-Poisson system,
\begin{align}
  &v\pdv{f_\ii}{x} -\pdv{\phi}{x}\pdv{f_\ii}{v}=0
  \label{eq:Vlasov-phi}\\
  &\pdv[2]{\phi}{x}=n_{\ee}-Zn_{\ii}\equiv-\rho,
  \label{eq:poisson}
\end{align}
where $n_{\ii}=\int_{-\infty}^{\infty} f_{\ii}\,\dd{v}$.
To keep the following discussion focused on collisional effects and
avoid issues with inter-species collisions, the model considered in
this paper only concerns single ion species distributions,
$f_\ii\to f$, and the electrons are assumed to be Maxwell-Boltzmann
distributed, which results in $n_{\ee}=n_{\ee,1}\exp(\phi)$, where
$n_{\ee,1}=Zn_{\ii}(\phi\to0)$ is the electron density which balances
both the incoming and reflected ions. Note that $n_{\ee,1}\neq1$, because
the normalized electron density only takes the value $1$ in the
\emph{unperturbed} upstream plasma, i.e.\ where the reflected ion beam
is not yet present.

The solution to \eqref{eq:Vlasov-phi} is given by $f(x,v)=f(\En)$,
where $\En=v^2/2+\phi(x)$ is the total ion energy; that is, ions
stream along constant-energy contours in phase space, see
figure~\ref{fig:phasespace}b.  With the assumption that the incoming
ions have a Maxwell-Boltzmann distribution, the ion distribution
function in regions~I and IV
becomes\footnote{ The factor $1/Z$ in the distribution function is due
  to the normalization of $\nn=1$, together with quasi-neutrality
  considerations in the far upstream of the shock. Furthermore, in our
  normalization, $Z$ does not enter the calculation in this single ion
  species problem, as the $Z$ values in \eqref{eq:poisson} and in
  \eqref{eq:fIIV} cancel, and the remaining $Z$ dependence is absorbed
  into our definition of $\tau=ZT_{\ee}/T_{\ii}$. }
\begin{equation}
f^{\rm I,IV}=\frac{1}{Z}\sqrt{\frac{\tau}{2\upi}}
\exp[-\frac{\tau}{2} \qty(\sqrt{v^2+2\phi} -\Ma )^2],
\label{eq:fIIV}
\end{equation}
where $\tau\equiv ZT_{\ee}/T_{\ii}$. We also divide up phase
space into four different regions: passing, trapped, co-passing, and
reflected regions of phase space, which we denote by the roman
numerals I, II, III, and IV, respectively. In the collisionless case,
regions II and III are completely empty, which consequently means
that the ion distribution is discontinuous at the separatrix. The
separatrix is marked out by the dashed line in
figure~\ref{fig:phasespace}b, and it is given by 
$\pm v_0=\pm\sqrt{2(\phimax-\phi)}$ in the up- ($+v_0$) and downstream
($-v_0$). It is this discontinuity of the distribution function to
which we will turn our attention in the following sections.

\subsection{Introducing collisions}
\label{sec:assumptions}
In the following, we will describe the underlying assumptions of the
collisional shock model. While the collisionless model has a
steady-state solution with a discontinuity, that discontinuity
can clearly not survive in the collisional problem. In this paper, we
consider a model problem where this discontinuous $f$ is taken as the
initial condition for the collisional problem, which has a time
dependence resulting from collisions. We assume that the collision
frequency is small; in particular, the collision time is much longer
than the typical time for ions to stream through some finite vicinity
of the shock front considered, i.e.\ across a few downstream
oscillations.

\begin{figure}
\centering
\includegraphics[width=0.5\columnwidth]{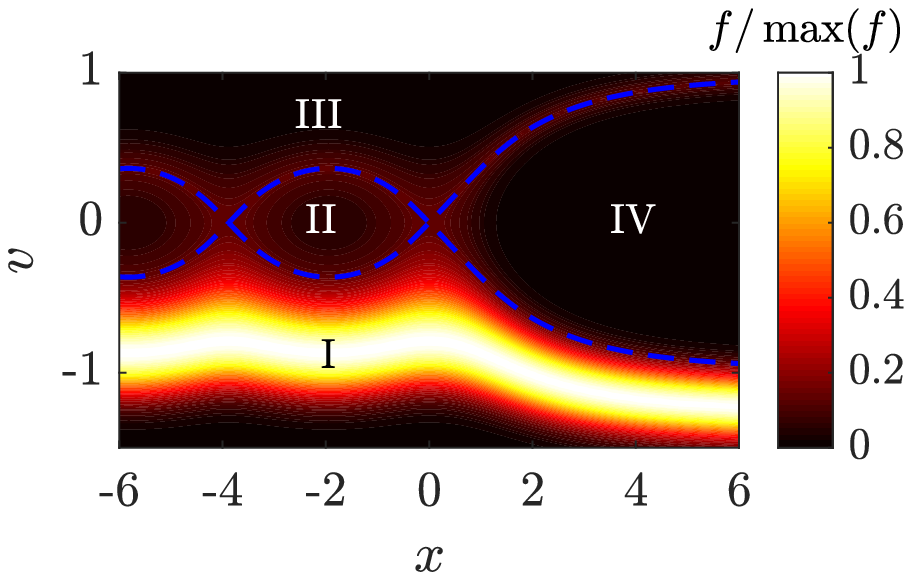}
\put(-167,97){\large \textcolor{white}{(a)}}
\hspace{1ex}
\includegraphics[width=0.3\columnwidth]{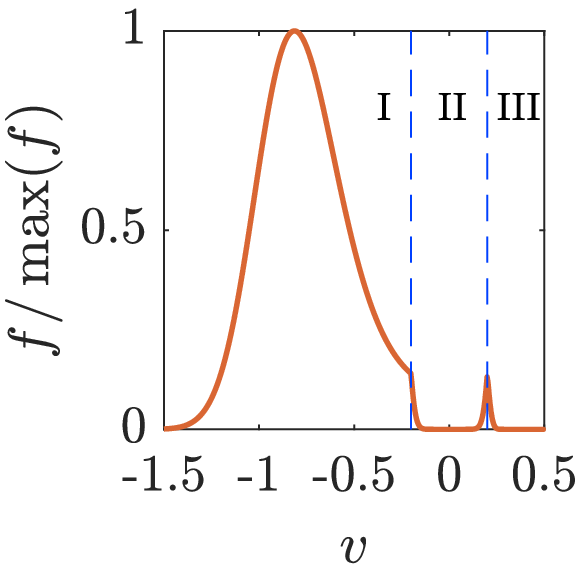}
\put(-83,96){\large (b)}
\centering
\caption{(a): Density plot of the ion distribution function at a
  collisional age of $\nus t=0.01$, for the parameters $\tau=50$,
  $\Ma=1.25$ (calculated using the model of
    section~\ref{sec:pert}). Dashed curves: phase-space separatrices. 
  Regions of phase space: I -- passing, II -- trapped,
  III -- co-passing, IV -- reflected.
  (b): The ion velocity distribution at $x=-0.62$, showing
  the counter and co-propagating populations of the trapped ions
  (region II).
}
\label{fig:distfunc}
\end{figure}
    
The collisions will, heuristically, act to smooth out the original
discontinuity into a thin boundary layer. As ions are scattered into
this boundary layer, they enter the trapped region (II), where they
will orbit; a population of trapped ions will develop along both the
upper and lower separatrices, as is illustrated in
figure~\ref{fig:distfunc}a. A cut of the same distribution at a
certain downstream location is shown  in figure~\ref{fig:distfunc}b.
There will also be scattering of ions near the upper separatrix into
region III. 

Far from the boundary layers, where the distribution function is not
as sharp, the solution remains essentially unchanged. Thus, for
$x\ge0$, we may assume $f$ to be the collisionless solution in regions
I and IV.  There is, however, a boundary layer in region I, which is
due to the depletion of ions near the separatrix in region I, but
since this layer is very thin and almost static in time, we disregard
it and focus on the thicker (but still thin compared to the thermal
speed) and time varying boundary layer which develops in regions II
and III.  We therefore use the value of $f^\mathrm{I,IV}$, from
\eqref{eq:fIIV}, on the separatrix, $v=-v_0$, as a boundary condition
for $f$ in region II.

For this problem, we would like to solve the ion kinetic equation,
which in the 1D electrostatic case reads
\begin{equation}
\pdv{f}{t}+v\pdv{f}{x}-\pdv{\phi}{x}\pdv{f}{v}=C[f],
\label{eq:fpe1} 
\end{equation}
where $C$ is the collision operator.  Since we consider the dynamics
of a thin boundary layer, we only need to focus on the highest-order
derivative term -- the diffusion.  Later, as we will introduce other
phase-space coordinates, first-order derivatives will be
systematically neglected, which will be justified
\emph{a~posteriori}. For simplicity, we will also neglect the velocity
dependence of the collision frequency.  With these choices, it is
sufficient to replace $C[f]$ by $(\nus/\tau)\p_{vv}$, where the
collisionality is assumed to be small, $\nus\ll 1$.  We have chosen to
define the collisionality, $\nus=\nu(v_{\ii,\text{th}})\lD/\cs$, using
the natural time normalization, $\lD/\cs$, and the collision frequency
at the ion thermal velocity, $v_{\ii,\text{th}}/\cs=\tau^{-1/2}$. This choice
is the reason for the factor $1/\tau$ in our model collision operator.

In our analytical model, we only consider the effects of collisions
within the narrow boundary layer around the separatrices. This
treatment of collisions effectively means that the collisionality
between high-energy ions and the bulk is suppressed. We point this
fact out since the shocks will have structures that are widely
separated in velocity space, e.g.\ the incoming and reflected
ions. Consequently, any attempt at simulating collisional shocks must
be done with a collision operator which accurately captures the
diminishing collisional effect on particles with high relative
velocities. In practice, this comment applies to any simulation with a
strongly super-thermal population.  We elaborate further on
collisional simulations in Appendix~\ref{sec:kinetic}, with a focus on
the Lenard-Bernstein model collision operator.

\subsection{Perturbative, orbit-averaged solution
  to the kinetic equation}
\label{sec:pert}

To solve \eqref{eq:fpe1}, we employ a perturbative scheme in the
ordering $\nus\ll1$, and we assume that all explicit time
dependence in the frame of the shock is due to collisions. As
such, all dependencies on $t$ vary on much slower time scales than
that of ions streaming past a few downstream oscillations. With
$f(x,v,t)$ perturbatively expanded as $f=f_0+f_1+\dots$, where
$f_{k+1}/f_{k}$ is small in $\nus$, the lowest order equation becomes
\begin{equation}
v\pdv{f_0}{x}-\pdv{\phi}{x}\pdv{f_0}{v}=0,
\label{eq:fpez} 
\end{equation}
which recovers \eqref{eq:Vlasov-phi} used for the collisionless
model. Analogously to the collisionless model, the solution to
\eqref{eq:fpez} is given here by $f_0(x,v,t)=f_0(\En,t)$, again with
$\En=v^2/2+\phi$. However, whereas the collisionless model was static
and the energy dependence was derived from a Maxwellian ion
distribution with the addition of reflected ions, the energy and time
dependence of $f_0$ is as yet undetermined; to obtain an equation for
that, we need to consider the next-order correction to the kinetic
equation,
\begin{equation}
\pdv{f_0}{t}+v\pdv{f_1}{x}-\pdv{\phi}{x}\pdv{f_1}{v}
=\frac{\nus}{\tau} \pdv[2]{f_0}{v}.
\label{eq:fpeo} 
\end{equation}
In analogy to gyrokinetics, a closed equation for $f_0$ is obtained by
taking an appropriate orbit average of \eqref{eq:fpeo} that
annihilates the $f_1$ terms. We employ the following orbit average
\begin{equation}
\ev{g}_\En\equiv
\qty[\oint_\En \dd\theta]^{-1} \oint_\En g\,\dd\theta
\equiv \qty[\oint_\En \frac{\dd{x}}{v}]^{-1}
\oint_\En \frac{g \,\dd{x}}{v},
\label{eq:defaver} 
\end{equation}
where the integrals are taken along constant $\En$ contours, over a
bounce period for trapped particles, and over a full oscillation
period of the downstream potential for passing particles. For any $g$
that is periodic in these domains, we find that 
\begin{equation}
\ev{v\pdv{g}{x}-\pdv{\phi}{x}\pdv{g}{v}}_\En
\propto \oint_\En\pdv{g}{\theta}\,\dd\theta=0.
\label{eq:averprop} 
\end{equation} 

We have found from \eqref{eq:fpez} that $\dv*{f_0}{\theta}=0$ and we
assume $f_1$ to have the required periodicity to make the $f_1$ terms
in \eqref{eq:fpeo} vanish upon orbit averaging, which gives 
\begin{equation}
\pdv{f_0}{t}
\approx \frac{\nus}{\tau} \ev{v^2}_\En \pdv[2]{f_0}{\En},
\label{eq:fpeaveraged} 
\end{equation}
where we have neglected all first-order derivative $\p_\En f_0$ terms
against the second-order derivative term, due to the sharp variation of
$f_0$ across the boundary layer, and used that $\p_{\En\En}f_0$ is
$\theta$-independent to pull it through the orbit average. 
We have hereby obtained an equation with only $f_0$, which can
be used to solve for the energy and time dependence of $f_0$. 

In order to explicitly evaluate $\ev{v^2}_\En$, we need to specify
$\phi(x)$. Orbit averages defined by \eqref{eq:defaver} for an
arbitrary $\phi(x)$ would not lead to closed form expressions. To find
the qualitative behaviour while keeping the problem analytically
tractable, we assume a simple harmonic oscillation of the downstream
potential, which can be justified in the low amplitude limit, where
the downstream oscillations reduce to linear ion acoustic
oscillations. We thus write
\begin{equation}
\phi(x)=\phimin+\phia \sin^2\left(\frac{\upi x}{\lambda}\right),
\label{eq:phiform} 
\end{equation}
where $\lambda$ is the wavelength of the downstream oscillation, and
$\phia=\phimax-\phimin$ with $\phimax$ and $\phimin$ the maximum and
minimum downstream values of $\phi$, respectively. Note that while
normally $x=0$ denotes the location of the first potential maximum, as
in figure~\ref{fig:phasespace}, when evaluating orbit averages we set
$x=0$ at a downstream potential minimum.

We also introduce $k=\sqrt{(\En-\phimin)/\phia}$, for which $k<1$ in
the trapped regions and $k\ge 1$ outside. The integrals of the
orbit average in the passing region can now be evaluated to:
\begin{equation}\label{eq:passingav}
\begin{aligned}
&\oint_\En \dd\theta
= \frac{1}{\sqrt{2}}\oint\frac{\dd{x}}{\sqrt{\En-\phi(x)}}
=\frac{2 \lambda}{\sqrt{2\phia}\upi k}
\int_0^{\upi/2}\frac{\dd{y}}{\sqrt{1-k^{-2}\sin^2y}}
=\frac{\sqrt{2}\lambda}{\upi\sqrt{\phia} k}K(k^{-2}),\\
&\oint_\En v^2\,\dd\theta
= \sqrt{2}\oint\sqrt{\En-\phi(x)}\,\dd{x}
= \frac{2\sqrt{2\phia} k \lambda }{\upi}
\int_0^{\upi/2}\sqrt{1-k^{-2}\sin^2y}\,\dd{y}
=\frac{2\sqrt{2\phia} k\lambda}{\upi}E(k^{-2}),
\end{aligned}
\end{equation}
where $K$ and $E$ denote the complete elliptic integrals of the first
and second kind, $y=\upi x/\lambda$, and we made use of the symmetry of
the potential about the potential minimum. Thus, we have
\begin{equation}
\eval{\ev{v^2}_\En}_{k\ge 1}=2\phia
k^2\frac{E(k^{-2})}{K(k^{-2})}.
\label{eq:passingaverage} 
\end{equation}
In the trapped region, the calculation is slightly more complicated,
since the particle does not sample the entire
$[-\lambda/2,+\lambda/2]$ region, only
$[-\lambda_\En/2,+\lambda_\En/2]$, where the limits
$\pm\lambda_\En/2$ depend on $\En$. Introducing $z$ such that
$k \sin z=\sin y$, we find
\begin{equation} \label{eq:trappedav}
\begin{aligned}
&\oint_\En \dd\theta
=\frac{2}{\sqrt{2}}\int_{-\lambda_\En/2}^{\lambda_\En/2}
\frac{\dd{x}}{\sqrt{\En-\phi(x)}}
=\frac{2 \sqrt{2}\lambda}{\sqrt{\phia}\upi k}
\int_0^{\upi/2}\frac{\dd{z}}{\sqrt{1-k^2\sin^2z}}
=\frac{2\sqrt{2}\lambda}{\upi\sqrt{\phia}}K(k^{2}),
\\
&\oint_\En v^2 \,\dd\theta
= 2\sqrt{2}\int_{-\lambda_\En/2}^{\lambda_\En/2} \sqrt{\En-\phi(x)}\,\dd{x}
=\frac{4\sqrt{2\phia}\lambda}{\upi}\left[(k^2-1)K(k^2)+E(k^2)\right],
\end{aligned}
\end{equation}
which yields
\begin{equation}
\eval{\ev{v^2}_\En}_{k<1}
=2\phia \qty[\frac{E(k^2)}{K(k^2)}-1+k^2].
\label{eq:trappedaverage} 
\end{equation}
With the explicitly evaluated orbit averages,
\eqref{eq:passingaverage} and \eqref{eq:trappedaverage}, we can now
express \eqref{eq:fpeaveraged} as
\begin{equation}
\pdv{f_0}{t}=2\frac{\nus}{\tau}\phia k^2 F(k^2)
\pdv[2]{f_0}{\En},\quad
F(k)=
\begin{cases}
\frac{1}{k^2}\left[ \frac{E(k^2)}{K(k^2)}+k^2-1 \right]
&\text{for }k<1, \\
\frac{E(k^{-2})}{K(k^{-2})} &\text{for }k\ge 1.
\end{cases}
\label{eq:fpeexpl1} 
\end{equation}

\subsection{Transformation to a diffusion equation and solution} 
\label{sec:diff}
It is practical to rewrite \eqref{eq:fpeexpl1} as a simple diffusion
equation
\begin{equation}
\pdv{f_0}{t}\approx\frac{\nus}{2\phia\tau}F(k)\pdv[2]{f_0}{k}
=\frac{F(\eps)}{\Upsilon^2}\pdv[2]{f_0}{\eps}\approx
\pdv[2]{f_0}{w}, 
\label{eq:fpeexpl2} 
\end{equation}
where the approximations made are that only second-order derivatives
are kept. The intermediary variable, $\eps=1-k$, is positive in the
trapped region and negative outside, and $w$ is defined by
\begin{equation}
\dv{w}{\eps}=\frac{\Upsilon}{\sqrt{F(\eps)}},\qquad
w(\eps=0)=0,\qquad \Upsilon=\sqrt{2 \phia \tau/\nus}.
\label{eq:wepstransf} 
\end{equation}
Thus, the stretched variable $w$ is defined to be order unity across
the boundary layer, while $\eps\ll 1$. Henceforth, the zero subscript
of $f$ is dropped to streamline notation, as $f_1$ is unimportant to
this order in $\nus$.

We use the collisionless distribution function, \eqref{eq:fIIV}, as
both the initial condition for this problem as well as the boundary
condition on the separatrix between regions I and II, assuming that the
collisionality is low enough that the shock has time to form on a much
faster time scale. In region II, $f$ is solved for using  
\begin{equation}
\pdv{f}{t}=\pdv[2]{f}{w}, 
\label{eq:fpeexpl3} 
\end{equation}
where we take the value of $f^{\rm I}$ at the separatrix 
\begin{equation}
\Fm\equiv f^{\rm I}(v{=}{-}v_0)
=\frac{1}{Z}\sqrt{\frac{\tau}{2\upi}}
\exp[ -\frac{\tau}{2}\qty(\sqrt{2\phimax} -\Ma)^2]
\label{eq:fI}
\end{equation}
as a boundary condition, and $v_0=\sqrt{2(\phimax-\phi)}$ is the
magnitude of the velocity at the separatrix. We do this since region I
is constantly replenished with new ions passing from the upstream.

The solution to the diffusion equation \eqref{eq:fpeexpl3} with the
initial condition $f(t{=}0,w)=\Theta(-w)$, where $\Theta$ denotes the
Heaviside step function, together with the boundary conditions
$f(w{\to}{-}\infty)=1$ and $f(w{\to}{+}\infty) =0$, is
$f(w,t)=\frac{1}{2}\erfc[w/(2\sqrt{t})]$, where $\erfc$ is the
complementary error function. This is easily shown employing the
Green's function $G(t, w-w')=(4\upi t)^{-1/2}\exp[-(w-w')^2/(4t)]$.
Noting that for this $f(w,t)$, $f(w{=}0)=\frac{1}{2}$ for all times,
it is clear that $f$ can also be considered as the solution for the
problem in the semi-infinite domain, where the boundary conditions are
$f(w{=}0)=\frac{1}{2}$, $f(w{\to}{+}\infty)=0$, and the initial
condition is $f(w{>}0,t{=}0)=0$. The boundary condition
$f(w{=}0)=\frac{1}{2}$ can only be sustained by a net influx of
particles across the boundary. The particles streaming along the lower
separatrix, coming from the upstream and keeping $f$ fixed in
region~I, represent a reservoir that provides this time-dependent
influx into region II (with the slight difference that the value of
$f$ at the separatrix is $\Fm$ instead of $\frac{1}{2}$). The
time scale separation between the streaming and the collisional
processes guarantees that $f$ is held fixed at the boundary, even
though there is a net outflux to region III, on top of the random
walk of particles away from the separatrix, deeper into II. The fact
that, formally, $w$ only has a finite range in region II is not a
concern, since this range is large for a small value of our
perturbation parameter $\nus$, and $\erfc$ drops rapidly for an
argument larger than unity. Hence the solution is
\begin{equation}
f^{\rm II,III}=\Fm \erfc\left(\frac{\pm w}{2\sqrt{t}}\right),
\label{eq:f23}
\end{equation}
where the sign of the argument in region II (III) is $+$ ($-$). Note
that there is some ambiguity of $f$ in region III, as it depends on
the behaviour of $f$ far downstream. This solution assumes an infinite
downstream oscillation, and as such, it represents an upper bound to
the density contribution from region III. We will find that the actual
behaviour of $f$ in region III has only a minor effect: $f$ being
finite in this region leads to a slight reduction of $\phimax$.

The resulting densities in regions II and III, due to (\ref{eq:f23}), are
\begin{equation}\label{eq:dens}
\begin{aligned}
n^{\rm II}=& 2\int_{0}^{v_0} f^{\rm II}(v) \,\dd{v}
=2\Fm\sqrt{2\phia}\int_{\kappa}^{1}\erfc\qty(\frac{w(k)}{2\sqrt{t}})
\frac{k\, \dd{k}}{\sqrt{k^2-\kappa^2}},
\\ n^{\rm III}=& \int_{1}^{\infty} f^{\rm II}(v)\, \dd{v}
=\Fm \sqrt{2\phia}\int_{1}^{\infty}\erfc\qty(-\frac{w(k)}{2\sqrt{t}})
\frac{k\, \dd{k}}{\sqrt{k^2-\kappa^2}},
\end{aligned}
\end{equation}
where $\kappa=\kappa(\phi)=\sqrt{(\phi-\phimin)/\phia}$; in region~II,
we used that $f$ is even in $v$. The practical aspects of how these
integrals are evaluated, along with further details on the numerical
implementation of the model, are discussed in
Appendix~\ref{sec:appnum}.  These densities, together with the
velocity integrals of $f^{\rm I}$ and $f^{\rm IV}$ ($v$ from $-\infty$
to $-v_0$ and $v_0$, respectively) given by \eqref{eq:fIIV}, are used
in Poisson's equation \eqref{eq:poisson} to calculate the potential. 

Since the ion distribution function implicitly depends on $\phimax$
and $\phimin$, these need to be calculated before Poisson's equation
can be integrated to obtain $\phi(x)$. This is achieved by introducing the
Sagdeev potential, $\Phi(\phi;\phimax,\phimin)$, with the property
$\dv*{\Phi}{\phi}=-\dv*[2]{\phi}{x}$ \citep{tidmankrall}. Thus,
\begin{equation}\label{eq:PHIu}
\Phi^{\rm u}(\phi;\phimax,\phimin)
=\int_{0}^{\phi}\rho^{\rm u}(\phi';\phimax,\phimin)\,\dd\phi'
\end{equation}
in the upstream and
\begin{equation}\label{eq:PHId}
\Phi^{\rm d} (\phi;\phimax,\phimin) =
\int_{\phimax}^{\phi}\rho^{\rm d}(\phi';\phimax,\phimin) \,\dd\phi'
\end{equation}
in the downstream region. Then the potential extrema can be found
solving the system
$\Phi^{\rm u}(\phi{=}\phimax;\phimax,\phimin)=0$ and
$\Phi^{\rm d}(\phi{=}\phimin;\phimax,\phimin)=0$ simultaneously. Note
that $\rho^{\rm u}$ and $\rho^{\rm d}$ are slightly different due to
the reflected ions, which indeed makes \eqref{eq:PHIu} and
\eqref{eq:PHId} two independent equations for the two unknowns
$\phimax$ and $\phimin$. This completes the calculation of $\phi(x)$
for any given instance of time.

Before we start discussing the results, we revisit the practice of
neglecting first-order derivatives across the boundary layer. If the
first-order energy derivative was not neglected on the right-hand side
of \eqref{eq:fpeaveraged}, we would get
\begin{equation}
\ev{\pdv[2]{f_0}{v}}_\En
=\ev{v^2}_\En \pdv[2]{f_0}{\En}+\pdv{f_0}{\En}
=2 \phia k^2 F(k)\pdv[2]{f_0}{\En}+\pdv{f_0}{\En}.
\label{eq:noapprox}
\end{equation} 
The approximation to neglect the second term must break down in a
certain vicinity of the separatrix, since $\ev{v^2}_\En$ vanishes at
the separatrix.  To estimate the size of this region, we balance the
sizes of the two terms 
\begin{equation}
F(\delta\epsilon)\frac{f_0}{(\rmDelta\En)^2}
\sim\frac{f_0}{\rmDelta\En},
\label{eq:balance}
\end{equation} 
where $\rmDelta$ refers to the size of the collisional boundary layer
and $\delta$ to the size of the layer where the approximation breaks
down.  The width of the collisional layer in $\eps$ is
$\rmDelta\eps\sim\sqrt{t}/\Upsilon\sim\sqrt{\nus t}\ll1$, since the
width in $w$ is ${\sim}\sqrt{t}$, and although $\tau$ is usually
large, it is considered to be an order-unity quantity in our
perturbation theory, as is $\phia$. Thus, we also have
$\rmDelta\En=2\phia k\delta{k}\sim\sqrt{\nus t}$. Using the asymptotic
behaviour of $F(\eps)\simeq 2/\ln (8/|\eps|)$ for $\eps\rightarrow 0$,
\eqref{eq:balance} yields
\begin{equation}
-\frac{1}{\ln (|\delta\eps|)}\sim \rmDelta\En \sim \sqrt{\nus t}.
\label{eq:balance2}
\end{equation} 
We therefore find that the layer where the approximation breaks down
is exponentially small
\begin{equation}
\delta\eps\sim\exp(-\frac{1}{\sqrt{\nus t}})\ll\rmDelta\eps,
\label{eq:balance3}
\end{equation}
and the accumulated contributions to $f_0$ form first-order
derivatives are thus negligible.

As it is relevant to the present discussion, we note again that we
have also neglected another small boundary layer around the
separatrix of a time-independent width in $v$ of
${\sim}\sqrt{\nus}\ll1$.

To avoid significantly increasing the mathematical complexity of the
problem, in deriving \eqref{eq:fpeaveraged}, we have also neglected
the term
$\ev*{\dot{\En}\partial_\En{f_0}}_\En
=\ev{\partial_t{\phi}}_\En\partial_\En{f_0}$,
stemming from the coordinate transformation
$\{x,v\}\to\{\theta,\En\}$. Due to the $\sqrt{\nus t}$ time dependence
of the downstream potential oscillations -- as we shall find in
section~\ref{sec:results} -- this term is formally of the same order as
the collisional diffusion term that we keep. Physically, it describes
adiabatic trapping of ions as their trapping region grows in time, and
it speeds up the accumulation of ions in the trapping region. Although
not changing the character of the solution, and importantly the
$\propto \sqrt{t}$ dependence of the potential variation (as confirmed
by numerical solutions of the problem), it leads to a somewhat higher
effective diffusion rate. Accordingly, our results represent a lower
bound on the effect of collisions.

\section{Results}
\label{sec:results}

\begin{figure}
\centering
\includegraphics[width=0.7\columnwidth]{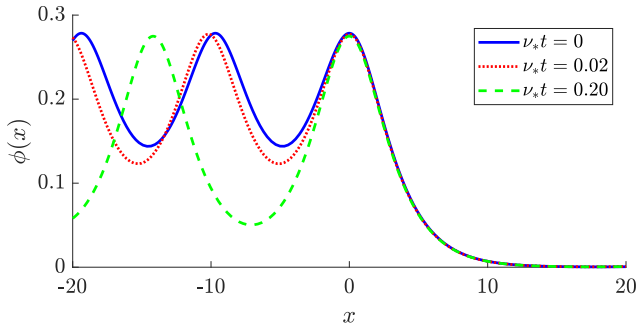}
\centering
    \caption{The variation of electrostatic potential over time due to
      collisions, for $\tau=50$, $\Ma=1.15$.  $\phi(x)$ is plotted for
      the collisional ages $\nus t=0$ (solid curve), $0.02$ (dotted),
      and $0.2$ (dashed).}
    \label{fig:phiintime}
\end{figure}

The qualitative effect of the collisional diffusion of ions is the
following: as the collisional boundary layer widens around the
phase-space separatrices with time, the difference between the
upstream and downstream densities decreases. Accordingly, the
downstream behaviour of the potential becomes increasingly similar to
that of the upstream, namely, the minimum of the potential $\phimin$
decreases, and the wavelength $\lambda$ increases. This behaviour is illustrated
in figure~\ref{fig:phiintime}, which shows the potential for three different
values of the \emph{collisional age}, $\nus t$. Unless the shock is
terminated by some other mechanism, this process would continue until
the width of the collisional layer becomes comparable to the ion
thermal width, at which point the shock has developed into a
symmetric, soliton-like structure. This stage of the evolution
corresponds to $\nus t\sim 1$; however, over that time scale, the
assumption of neglecting collisional friction as compared to diffusion
has already broken down. The maximum of the potential $\phimax$ is
less affected; it only changes due to the distribution function
becoming finite in region III. Indeed, if $f^{\rm III}$ would be set
to zero, $\phi_{\rm max}$ would stay constant in time, as it is only
affected by the upstream distribution function.

\begin{figure}
\centering
\includegraphics[width=0.4\columnwidth]{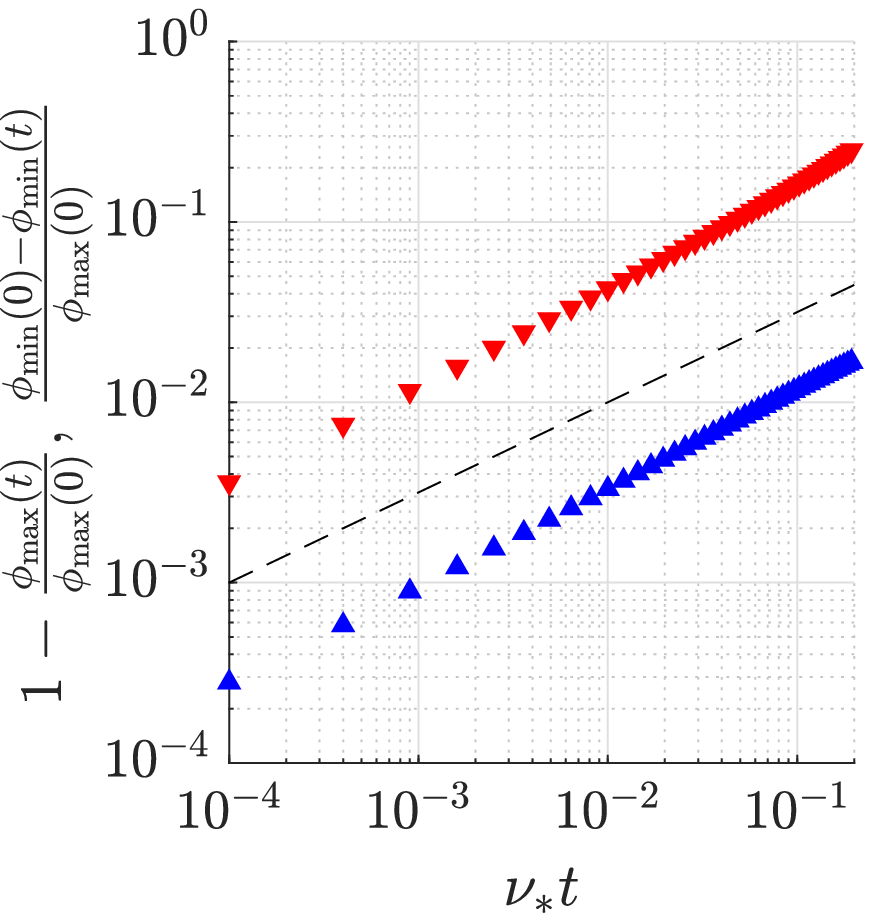}
\put(-30,130){\large (a)} 
\includegraphics[width=0.4\columnwidth]{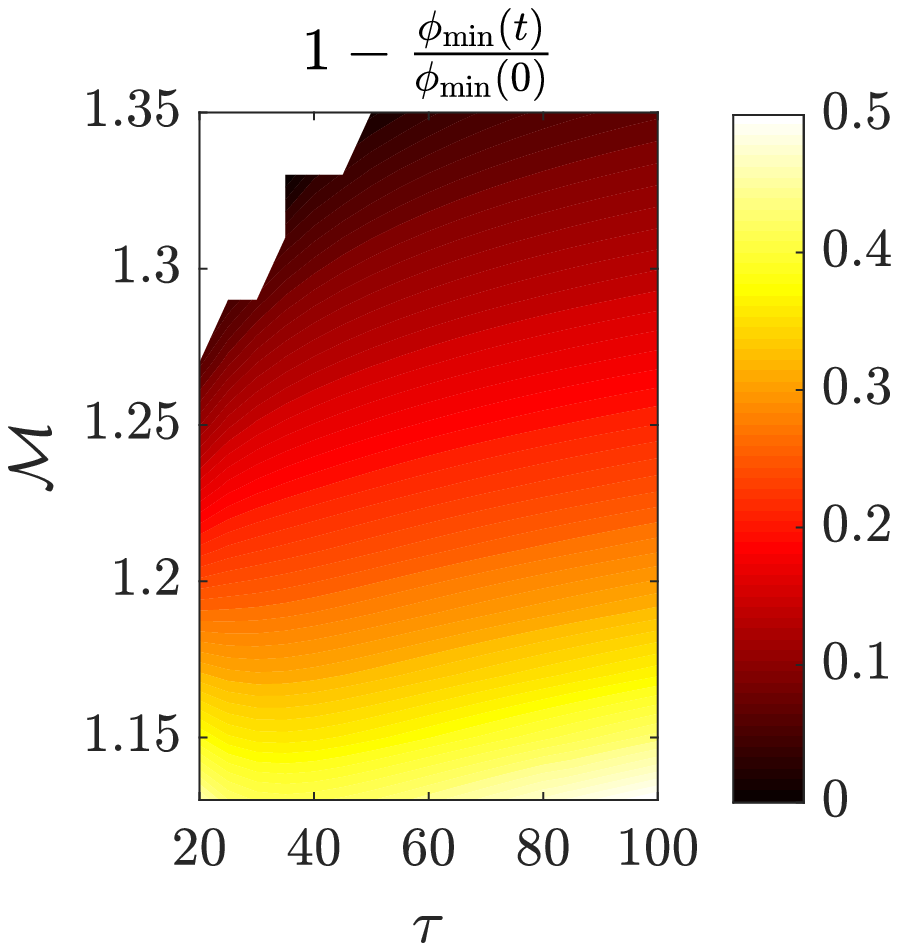}
\put(-60,130){\large \textcolor{white}{(b)}} 
\centering
    \caption{(a) Reduction in $\phimax/\phimax(t=0)$
      (\textcolor{blue}{$\blacktriangleup$}) and
      $\phimin/\phimax(t=0)$ (\textcolor{red}{$\blacktriangledown$})
      with collisional age, for $\tau=50$, and $\Ma=1.15$. For
      reference, the $\propto\sqrt{\nus t}$ dependence is indicated by
      the dashed line. (b) The $\Ma$ and $\tau$ dependence of
      $1-\phimin(t)/\phimin(t=0)$ at the collisional age
      $\nus t=0.1$. }
    \label{fig:single}
\end{figure}

The changes of both the potential maximum and minimum with collisional
age are both approximately proportional to $\sqrt{t}$, although the
effect on $\phimax$ is usually an order of magnitude smaller than that
on $\phimin$. This result is illustrated in the log-log plot of
figure~\ref{fig:single}a, showing $[\phimax (0)-\phimax (t)]/\phimax(0)$
and $[\phimin (0)-\phimin (t)]/\phimax(0)$ as functions of $\nus t$,
corresponding to the symbols $\blacktriangleup$ and
$\blacktriangledown$, respectively. The $\sqrt{t}$ dependence is
expected, since the width of the boundary layer is ${\sim}\sqrt{\nus
  t}$. However, the actual dependence is slightly stronger than
$\sqrt{t}$, and it is not strictly a power law, due to the mapping of
$w$ to $v$, and the nonlinear nature of the problem.

The importance of the collisional effects can be quantified by the
relative reduction of $\phimin$ for a given finite collisional
age. Figure~\ref{fig:single}b shows
$[\phimin(0)-\phimin(t)]/\phimin(0)$ at $\nus t=0.1$ as a function of
$\Ma$ and $\tau$. We have chosen the normalization such that the
solution would become soliton-like when this quantity would reach the
value $1$. Naturally, the importance of the collisions depends on the
size of the ion trapped region at $t=0$, and thus on the amplitude of
the downstream potential oscillation, characterized by $\phia$. There
is an upper limit in $\Ma$ for laminar electrostatic shock solutions
to exist, see e.g.\ figure~2 of \citep{CairnsPPCF}, and that is the reason
for the upper left corner of figure~\ref{fig:single}b being empty. At
this limit, $\phia$ reduces to zero; therefore, the effect of
collisions on $\phimin$ vanishes. Thus, the effectiveness of
collisions decreases with increasing $\Ma$, as seen in
figure~\ref{fig:single}b. The effect of collisions mostly increases with
$\tau$, for the same reason. Thus, for a fixed $\Ma$, a higher $\tau$
corresponds to a larger relative downstream oscillation.

\begin{figure}
\centering
\includegraphics[width=0.4\columnwidth]{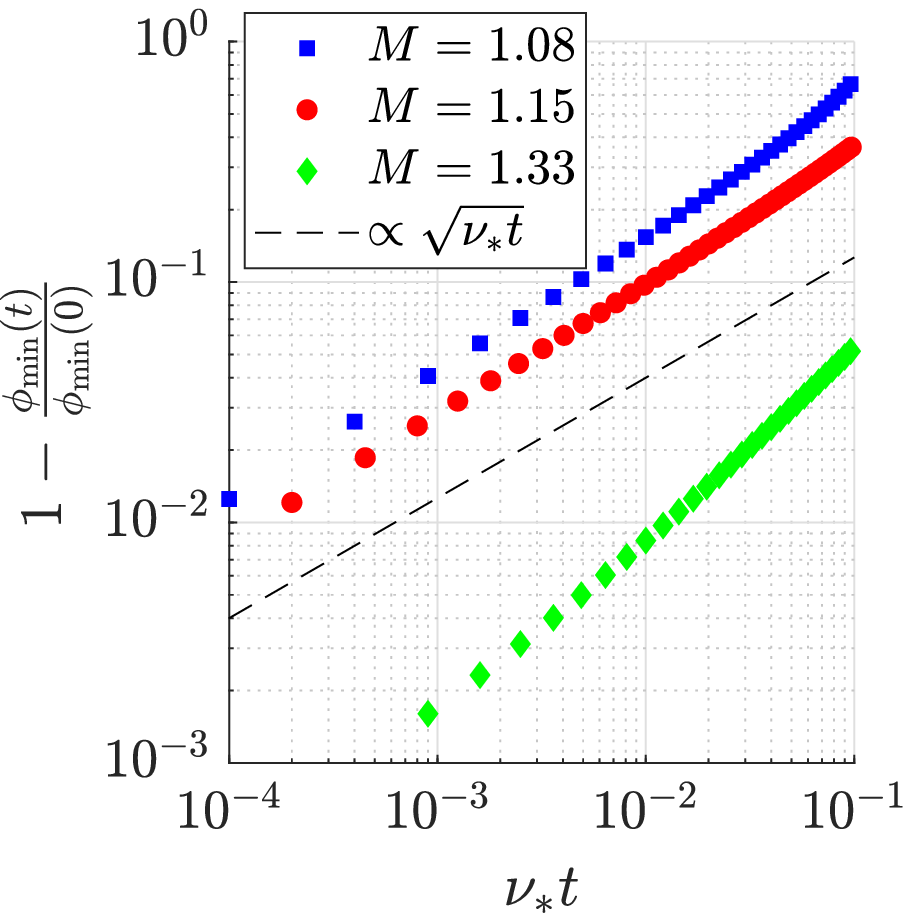}
\put(-30,45){\large (a)}
\includegraphics[width=0.4\columnwidth]{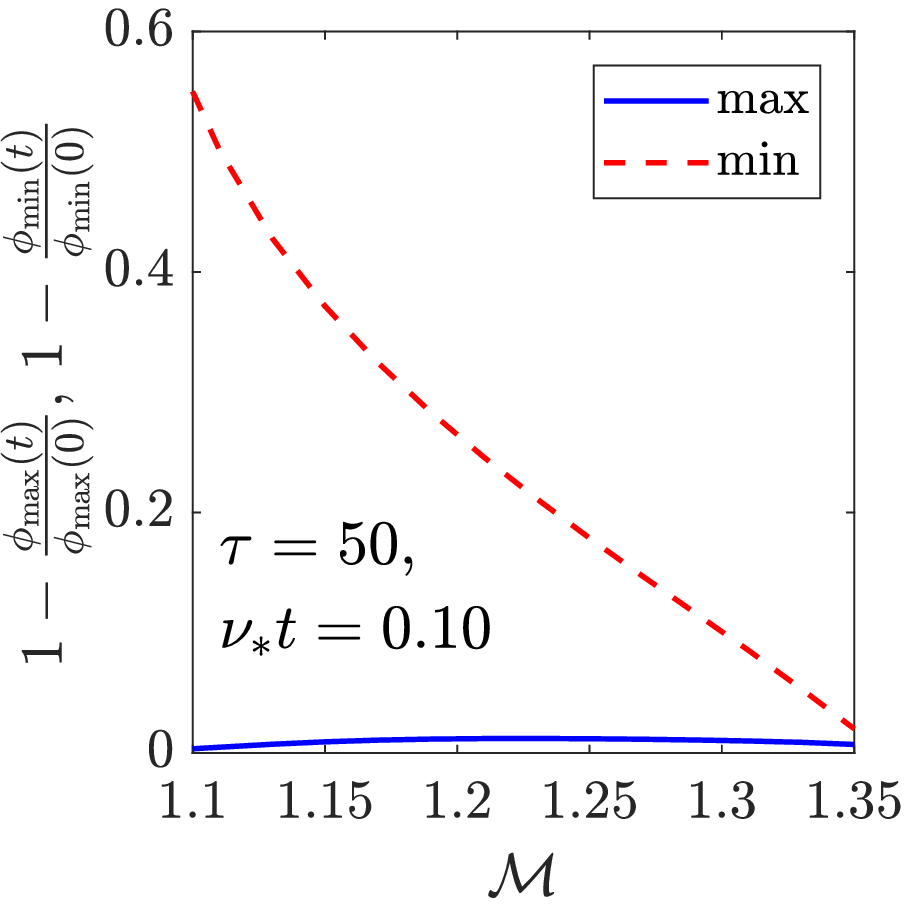}
\put(-30,45){\large (c)} \\
\includegraphics[width=0.4\columnwidth]{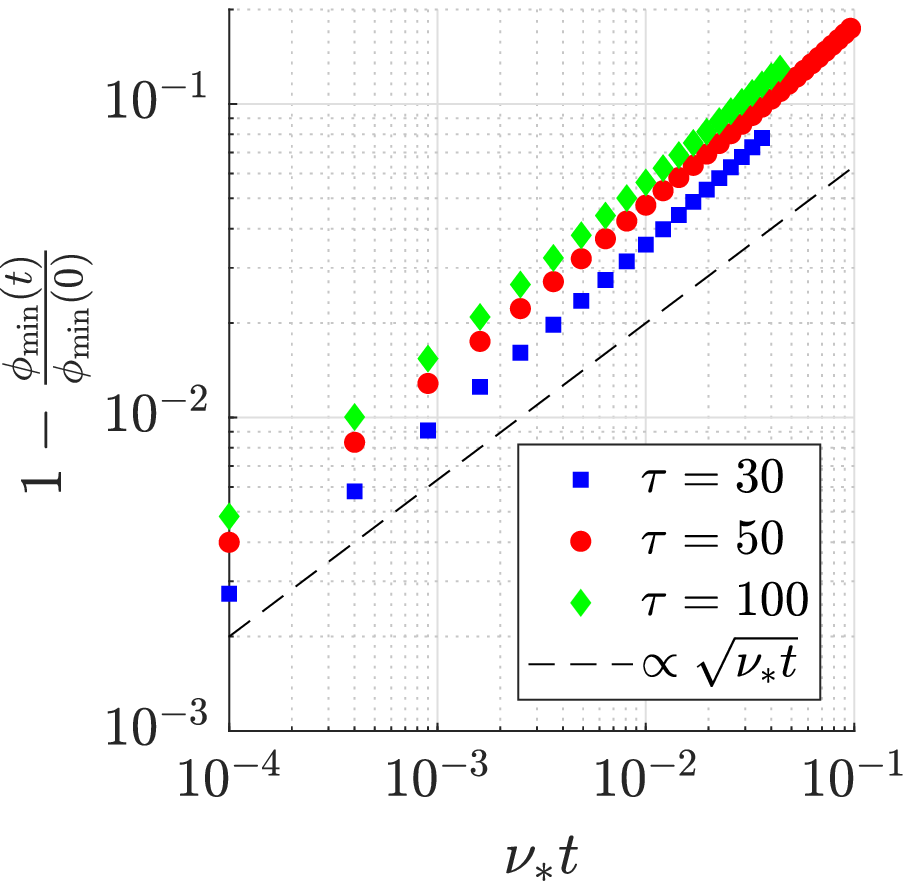}
\put(-110,130){\large (b)} 
\includegraphics[width=0.4\columnwidth]{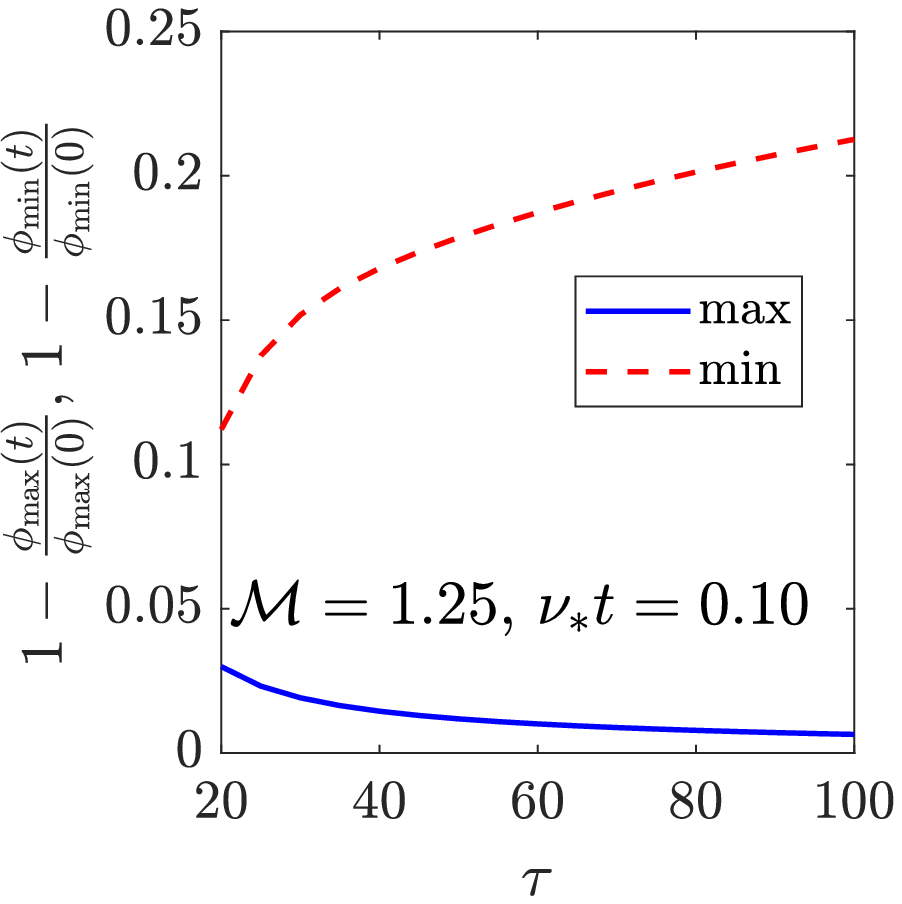}
\put(-110,130){\large (d)} 
\centering
\caption{(a-b): Relative reduction of $\phimin$ with collisional
  age. (a): For $\tau=50$; $\Ma=1.08$ (square symbols), $1.15$
  (circles), and $1.33$ (diamonds). (b): For $\Ma=1.25$; $\tau=30$
  (squares), $50$ (circles), and $100$ (diamonds).  For reference, the
  $\propto\sqrt{\nus t}$ dependence is indicated by dashed
  lines. (c-d) Relative reduction of $\phimin$ (dashed line) and
  $\phimax$ (solid), at $\nus t=0.1$. (c): $\Ma$ scan for
  $\tau=50$. (d): $\tau$ scan for $\Ma=1.25$. }
    \label{fig:scalings}
\end{figure}

The above parametric dependencies are analysed further in
figure~\ref{fig:scalings}. We find that the scaling of 
$[\phimin (0)-\phimin (t)]/\phimin(0)$ is close to the $\sqrt{\nus t}$
scaling observed for intermediate values of $\Ma$, as seen in
figure~\ref{fig:scalings}a. For high $\Ma$, where the trapped region of
the collisionless solution is small, we find an overall stronger
scaling. For low $\Ma$ the scaling gets stronger at larger collisional
ages when the effect of collisions on the potential becomes order
unity, and the downstream oscillation becomes significantly
non-sinusoidal. Besides showing the strong $\Ma$ dependence of the
collisional effects on $\phimin$, figure~\ref{fig:scalings}c also
illustrates that the effect on $\phimax$ remains negligible for all
$\Ma$ values. As seen in figure~\ref{fig:scalings}b, getting further
away from the limit of shock existence -- as $\tau$ increases -- also
leads to a scaling closer to $\sqrt{\nus t}$. We find that for lower
values of $\tau$, collisional effects on $\phimax$ increase, see
figure~\ref{fig:scalings}d. However, the corresponding changes on the
reflected ion fraction are still weak, as discussed in
Appendix~\ref{sec:apprefl}.

\section{Conclusions and discussion}
\label{sec:conc}
We have studied the effect of a weak but finite collisionality on the
dynamics of kinetic electrostatic shocks in one dimension. The electrostatic
reflection of ions results in the trapped and co-passing regions of
ion phase space being empty in the exact collisionless case. This
depletion of certain phase-space regions corresponds to a
discontinuity in the distribution function across the separatrix. In
the presence of collisions, ions are scattered into the originally
empty regions of phase space, leading to the development of
collisional boundary layers around the separatrix. To focus on this
process, we discuss only single ion species plasmas. We consider an
initial value problem, initialized by the solution to the
collisionless problem, and employ a perturbative, orbit-averaged
treatment based on the smallness of the ion collisionality
$\nus=\nu\lD/\cs$.

One might be tempted to neglect collisions when $\nus \ll 1$; however,
since particles keep accumulating in the ion trapped regions, the
important quantity for the process is not $\nus$, but rather the
collisional age, $\nus t$. Even though $\nus \ll 1$, the collisional
age can become substantial during the lifetime of the
shock. Furthermore, as expected from a diffusion problem, the width of
the collisional boundary layer is approximately proportional to
$\sqrt{\nus t}$. For the effect of collisions to be non-negligible,
low ion temperature and high electron density are required. Meanwhile,
the collisional effects depend only weakly on the electron
temperature. For a reference point, in a hydrogen plasma with 
$T_{\ii}=0.1\,\rm keV$ and $n_e=10^{27}\,\mathrm{m}^{-3}$,
$t\sim 30$ (in dimensional time,
$30\lD/\cs\sim0.7\,\mathrm{ns}$) corresponds to 
an order unity $\nus t$.

While the effect of collisions on the shock potential $\phimax$ is
small, as the trapped region gets populated by ions, the trapped
regions become increasingly similar to the reflected region close to
the shock. Accordingly, the minimum value of the downstream
electrostatic potential, $\phimin$, decreases towards the far upstream
value of $\phi$, which is zero in our choice of gauge, and the
wavelength of the downstream oscillations increases. These effects can
reach significant levels more rapidly when the trapped region in the
initial state is larger, corresponding to smaller values of $\Ma$ and
higher values of $\tau$. When $\nus t$ becomes order unity, our
mathematical model breaks down, but it is expected that the solution
becomes somewhat similar to a train of solitary waves, each one
becoming symmetric about its maximum.

We would also predict, from the rather weak effect collisions have on
$\phimax$ and the reflected ion population, that these types of
shocks have the potential to be used as relatively long time stable
high-energy ion sources -- compared to other laser time scales. Even
though the shock downstream might degenerate due to collisions, the
upstream stays rather unaffected, both in terms of number of reflected
ions and their energy. 

In more than one spatial dimensions ion-ion modes, beam-Weibel and
temperature anisotropy driven Weibel instabilities can become unstable
and represent limitations on the lifetime of the shock; see
\citet{Kato-Takabe2010} and references therein. Considering
scenarios where such instabilities develop and interfere with the
collisional process considered here is outside the scope of our
studies, but we presume that interesting synergistic effects from the
growing trapped regions in the downstream and anisotropy driven
instabilities may arise.

We have considered only the early development of electrostatic shocks
under the influence of collisions. Realizing the similarity between
the transport in phase space across trapped region and the transport
around magnetic islands in a magnetic confinement fusion device, the
long time asymptotic collisional behaviour of weakly collisional
shocks could be studied employing methods similar to those used by
\citet{hazeltine97}. Another possibility to generalize the
semi-analytical model of this paper would be to allow for multiple ion
species, where collisional friction between the various species can
become important \citep{Turrell2015}.


\acknowledgements
The authors are grateful for the fruitful
discussions with Ian Abel, Laurent Gremillet, Julien Ferri, T\"{u}nde
F\"{u}l\"{o}p, and Longqing Yi.
This work was supported by
the European Research Council (ERC-2014-CoG grant 647121),
the International Career Grant of Vetenskapsr{\aa}det (VR)
(Dnr.~330-2014-6313)
and Marie Sk{\l}odowska Curie Actions, Cofund, Project INCA 600398.
J. Juno was supported by a NASA Earth and Space Science Fellowship
(grant no. 80NSSC17K0428);
J.M. TenBarge was support by NSF SHINE award (AGS-1622306).
The development of the \gkyl{} code is partly funded by the
U.S. Department of Energy under Contract No. DE-AC02-09CH11466 and by
the Air Force Office of Scientific Research under grant number
FA9550-15-1-0193.
The simulations were performed on resources provided by the Swedish
National Infrastructure for Computing (SNIC) at Chalmers Centre for
Computational Science and Engineering (C$^3$SE).

\appendix

\section{Kinetic simulations,
  limitations of the Lenard-Bernstein operator}
\label{sec:kinetic}
Here, we first consider the collisional boundary layer formation in
initially discontinuous distributions in kinetic simulations, then we
discuss the limitations of the Lenard-Bernstein operator in the
presence of energetic populations. 

The development of a collisional boundary layer at a discontinuity of
the distribution function was reproduced in simulations with the
Vlasov-Maxwell solver in the open-source framework \gkyl{}\footnote{
\url{https://gkeyll.rtfd.io/}
(accessed: 2018-08-29)} \citep{juno2018}. 
While the implementation of a Fokker-Planck operator is ongoing,
collisions are currently modelled through a generalized
Lenard-Bernstein collision operator (LBO)
\citep{Lenard-Bernstein1958}, allowing a finite flow speed and
inter-species collisions (with some restrictions to avoid negative
entropy production). Here we only use the ion-ion collision part of
the operator that, in one dimension, reads
\begin{equation} \label{eq:LBO}
\CLBO[f]=\nus\pdv{v}\qty[(v-V)f+\vth^2\pdv{f}{v}],
\end{equation}
where $V$ and $\vth$ represent the flow and thermal speed of $f$ (the
latter defined such that it is $\sqrt{T/m}$ for a Maxwellian).

\begin{figure}
\centering
\includegraphics[width=0.35\columnwidth]{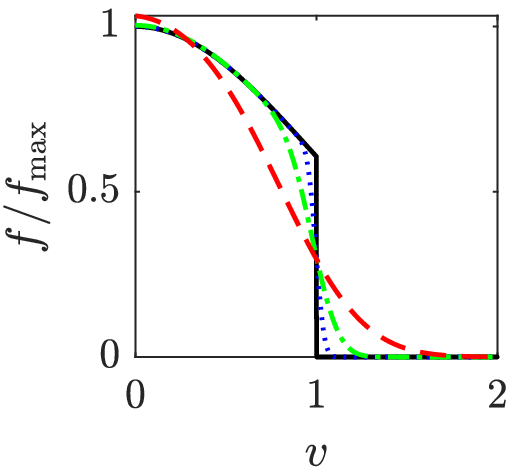}
\put(-30,98){\large (a)}
\includegraphics[width=0.35\columnwidth]{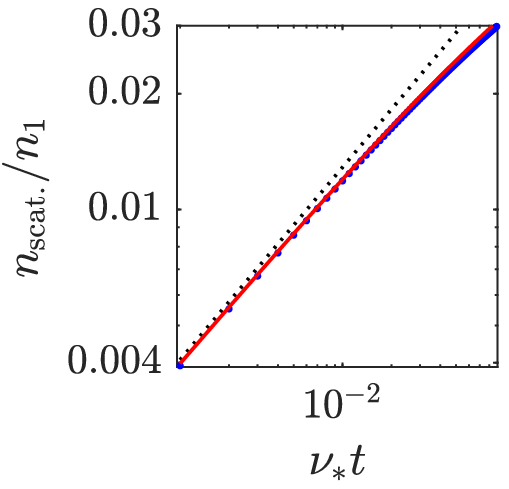}
\put(-90,98){\large (b)}
\caption{(a) Simulated distribution function near a sharp cutoff, in
  an otherwise Maxwellian ion distribution, at collisional ages
  $\nus t=0$ (black line), $10^{-3}$ (blue dotted), $10^{-2}$ (green
  dash-dotted), and $10^{-1}$ (red dashed). (b) The time evolution
  of the density of the ions which have been scattered out above the
  cutoff, simulated value (blue dots) compared to a theoretical
  estimate assuming only diffusion (red solid line), and the early
  time asymptotic behaviour $\propto \sqrt{t}$ (black dotted). } 
\label{fig:cut-maxwell}
\end{figure}

We first consider the time evolution of a spatially homogeneous
distribution that is Maxwellian with no flow and unit thermal speed for
$v<1$, and $0$ above. Note that this truncation means that $V<0$ and
$\vth<1$ for this distribution. The simulation uses 2048 cells in
velocity spanning $[-4 \cs,\,2 \cs]$, and a polynomial order of 2;
the electrons and the electric field were not evolved. The box is
$1\lD$ wide in configuration space, with 16 cells and periodic
boundary conditions. 

The sharp drop in $f$ is quickly smoothed out, as seen in
figure~\ref{fig:cut-maxwell}a, showing $f$ at four instances of time. In
figure~\ref{fig:cut-maxwell}b, the density of ions scattered above the
cutoff velocity, $n_{\text{scat.}}=\int_{1}^{\infty}f\,\dd{v}$, is
plotted against the collisional age $\nus t$. For very low
collisional age, the scattered density (blue dots) follows a
$\sqrt{\nus t}$ behaviour (black dotted line) very closely, but above
$\nus t \sim 0.01$ it starts to deviate, growing more slowly than
$\sqrt{\nus t}$. This behaviour is well approximated by taking the
integral $n_{\text{scat.}}^{(0)}=\int_{1}^{\infty}f_0\,\dd{v}$, where
\begin{equation}
f_0 = \frac{1}{\sqrt{2\upi}} \exp(-\frac{v^2}{2})
\frac{1}{2}\erfc\qty(\frac{v-1}{2\vth\sqrt{\nus t}}),
\end{equation}
which is plotted in the figure with red solid line.  At later times, as
the collisional drag becomes comparable with diffusion, the
distribution will eventually evolve towards a new Maxwellian
characterized by $V$ and $\vth$. However, the good agreement with the
theoretical estimate based on assuming a pure diffusion suggests that
neglecting drag is reasonable, even up to a collisional age of
${\sim}0.1$.

The Lenard-Bernstein operator (LBO) \citep{Lenard-Bernstein1958} is
often employed as a simple collision model that captures both
collisional diffusion and drag, and importantly, because it is
meaningful in one velocity dimension. By construction, the LBO has a
velocity independent collision frequency, which drives the
distribution towards a Maxwellian with density, flow speed, and
temperature determined by the first three velocity moments of the
original distribution function. This construction is to ensure the
conservation of particles, momentum, and energy.  However, the LBO is
not well suited for situations when the distribution has a significant
super-thermal population, as in our case.

The main contributor to the unphysical behaviour of the LBO is the
large effect super-thermal structures can have on the moments of the
distribution, especially on $\vth^2$. Take, for instance, the
collisional interaction between a high-energy beam and a bulk plasma
-- a situation which we have in the shock upstream. In reality (as in
the Fokker-Planck operator), the collisionality between the particles
in the beam and the bulk would decrease as $v^{-2}$, which would mean
that the collisional interaction between the beam and bulk would be
virtually non-existent.  However, the LBO has no such features, and
the high energy of the beam significantly skews the energy and
velocity moments; there will also be an artificially high drag due to
the linear increase of the friction term. Consequently, the bulk
plasma will also be noticeably affected, even when the beam only
constitutes a very low fraction of the total density. In the case of
our shock simulations, the reflected ions only made up approximately
1\,\% of the upstream plasma, but the artificially high collisional
effects still render the simulation unusable for testing our
analytical model.

\begin{figure}
\centering
\includegraphics[width=0.45\columnwidth]{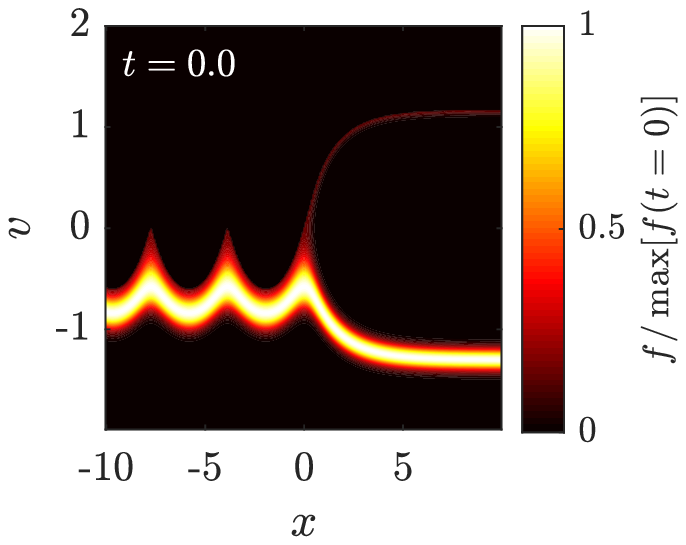}
\put(-142,37){\textcolor{white}{\large (a)}}
\hspace{2ex}
\includegraphics[width=0.45\columnwidth]{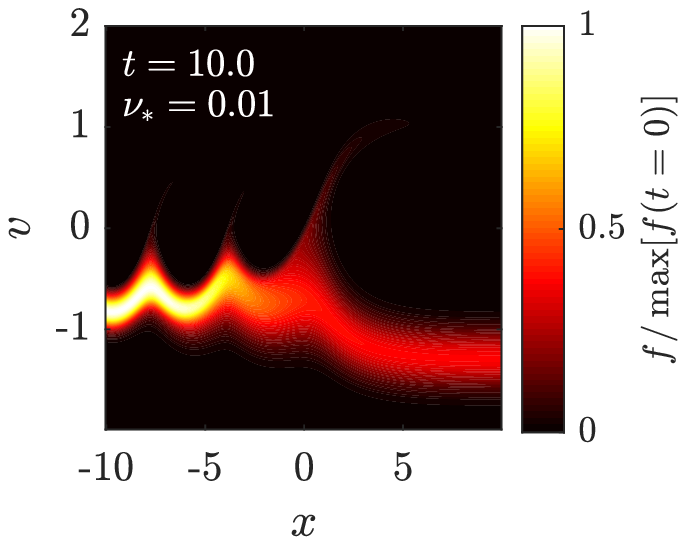}
\put(-142,37){\textcolor{white}{\large (c)}}
\\[2ex]
\includegraphics[width=0.45\columnwidth]{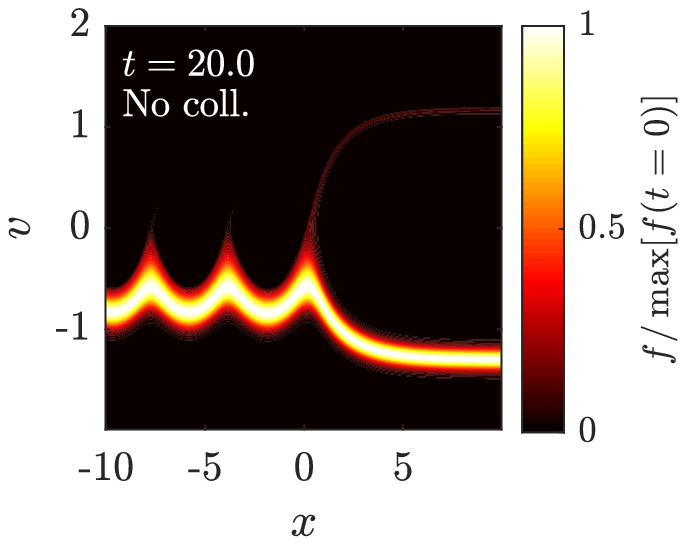}
\put(-142,37){\textcolor{white}{\large (b)}}
\hspace{2ex}
\includegraphics[width=0.45\columnwidth]{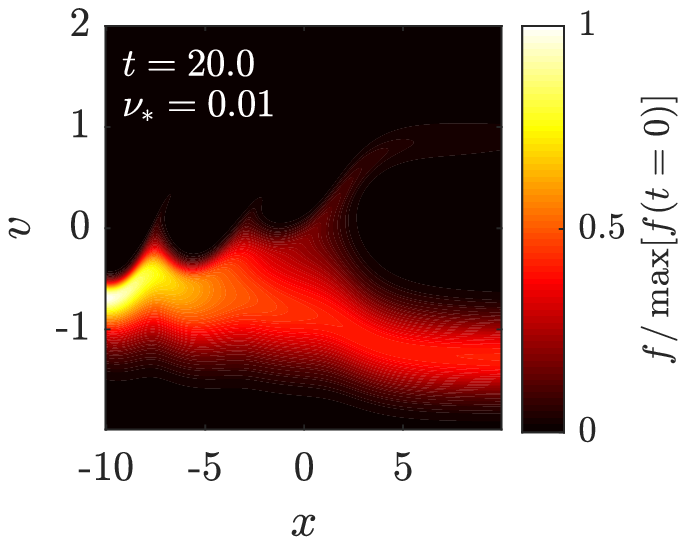}
\put(-142,37){\textcolor{white}{\large (d)}}
\caption{(a-d): Ion distribution functions at different stages of
  numerical simulations of shocks with $\Ma=1.3$ and $\tau=200$.
  (a): All simulations were initialized with the collisionless ion
  distribution function calculated from the analytical model.
  (b): The time-evolved distribution function with no collisions. This
  shows that the shock is static, as is required in the model. 
  (c-d): The time-evolved distribution function with the
  Lenard-Bernstein collision operator (LBO) acting on it. The
  unphysical collisional interaction between the high-energy reflected
  ions and the incoming bulk quickly destroys the shock structure. 
}
\label{fig:Sh-sims}
\end{figure}

To illustrate this problem, we have performed collisional simulations
of shocks in \gkyl{}, using the LBO. The shock simulations were
initiated with the collisionless shock ion distribution function. The
initial condition was constructed by calculating $\phi(x)$ from our
semi-analytical model, which was then used to initialize the ion
distribution function according to \eqref{eq:fIIV} in regions I and
IV. Above the separatrix, a rapid Gaussian cutoff was employed,
\begin{equation}
f(v{>}{\pm}v_0)
=f^{\rm I, IV}(\pm v_0)\exp[-10\frac{(v\mp v_0)^2}{\delta{v}^2}],
\end{equation}
where $\delta{v}$ is the simulation velocity grid size. This gradual
cutoff at the separatrix makes the simulation discretization
smoother, and thus also results in smoother simulation
results. Furthermore, the electrons were initialized to be
Maxwell-Boltzmann distributed with a flow velocity $-\Ma$, and the
initial electric field was simply taken as $-\dv*{\phi}{x}$. The
simulations are done in the shock frame, with $640$ cells in
configuration space covering $x=[-29,\,20]\lD$ and employ
perfectly matched layer boundary conditions. The velocity space
covers $v=[-3.5,\,3.5]\cs$ with $512$ cells for ions, and
$v=[-5,\,5]v_e=[-214,\,214]\cs$ with $256$ cells for electrons, with
the physical electron-to-proton mass ratio. A zero flux boundary
condition is used in velocity. The polynomial order is $2$.  

In figure~\ref{fig:Sh-sims}, we show the resulting ion distribution
functions at various simulation times (measured in $\lD/\cs$)
after the time evolution in \gkyl{}, both with (c-d) and without (b)
collisions in the simulation. The initial ion distribution function
for both simulation runs is shown in figure~\ref{fig:Sh-sims}a. From
figure~\ref{fig:Sh-sims}b, we see that the shock is indeed static, as
was expected and required by our semi-analytical model. The time
evolution to $t=20$, figure~\ref{fig:Sh-sims}b, required
${\sim}30\,000$ simulation time steps. This result demonstrates that the
initial condition is faithfully imported and shows the numerical
stability of the code.

However, the collisional simulation of the shock in
figure~\ref{fig:Sh-sims}c-d is clearly not static, and the collisional 
effects evolve at rates much faster than would have been expected from
the relatively low collisionality of $\nus=0.01$. We see that the
collisional effects mostly originate upstream ($x>0$) of the shock
(c), and then they eventually propagate downstream of the shock
(d). This result is precisely the expected behaviour of the LBO in a
scenario with a high-energy beam interacting with the bulk plasma: the
bulk heats up at an unphysical rate due to collisions with the
high-energy beam.

Since the LBO is relatively simple, it is often the first choice of
model collision operator to be implemented in a kinetic Eulerian
simulation code. In some circumstances, where $f$ remains close to a
Maxwellian, the above discussed artefacts of LBO will not arise. We
would, however, point out the risk of using the LBO in simulations
with more complex systems: 
Distributions which are strongly non-Maxwellian and/or have high-energy
structures will experience unphysical collisional effects due to
artificially strong collisionality between particles with a large
velocity separation.

\section{Numerical implementation of the analytical model}
\label{sec:appnum}
The solution for $f$, (\ref{eq:f23}), is given in terms of $w$, while
the densities are more conveniently written in terms of $k$,
(\ref{eq:dens}). Thus $w(k)$, or in practice $w(\eps)$, needs to be
evaluated. Note that the direct numerical integration of the equation
that relates them, (\ref{eq:wepstransf}), is problematic, since
$\dv*{w}{\eps}$ is divergent at $\eps=0$. For an accurate evaluation of
$w(\eps)$, the numerical integration of (\ref{eq:wepstransf}) can be
started at some finite $\eps$, where $w$ is given by its small $\eps$
asymptotic value
\begin{equation}
\frac{w(\eps)}{\Upsilon}=\frac{\eps}{\sqrt{2}}
\sqrt{\ln\left(\frac{8}{|\eps|}\right)}\pm 2\sqrt{2\upi}\,\erfc
\left[\sqrt{\ln\left(\frac{8}{|\eps|}\right)}\right]\qquad\text{as}\quad
\eps\rightarrow 0^{\pm}.
\label{asymptoticw}
\end{equation}
 
For a less accurate but significantly faster evaluation of $w(\eps)$,
we use an approximation function\footnote{
The scripts with these numerical implementations, together with the
other numerical tools developed for this paper, are freely available
at \url{https://github.com/andsunds/ShockLib} as open-source.}
$\hat{w}(\eps)$ that is defined in the following way. It takes the
asymptotic value \eqref{asymptoticw} for $0<\eps\le 0.1$, then for
$0.1<\eps\le 1$ it is given by
\begin{equation}
\frac{\hat{w}(\eps)}{\Upsilon}=\sqrt{2}(\eps-1)-
\frac{\sqrt{2}}{48}(1-\eps^3)+C^{+},
\label{asymptotic2}
\end{equation}
which is obtained by integrating the $k\rightarrow 0$ asymptote
\begin{equation}
\frac{1}{\Upsilon}\dv{w}{\eps}\simeq\sqrt{2}
\left(1+\frac{k^2}{16}\right),
\label{asymptotic3}
\end{equation}
and setting the constant of integration $C^{+}=1.4756$ by the
numerically determined value of $w(\eps=1)$. For negative values of
$\eps$ we use (\ref{asymptoticw}) for $-0.5\le\eps<0$, and for
$\eps\le -2$ we take
\begin{equation}
\frac{\hat{w}(\eps)}{\Upsilon}=\eps+C^{-},  
\label{asymptotic4}
\end{equation}
where in the $\eps\rightarrow -\infty$ limit, $\Upsilon^{-1}\dv*{w}{\eps}\simeq 1$
is integrated, and $C^{-}=-0.3310$ is determined by evaluating $w(\eps)$
at some sufficiently large negative value of $\eps$, namely
$w(\eps=-50)/\Upsilon=-50.3310$. For the region $-2\le\eps<-0.5$ a
linear curve matching to the two asymptotes is used. The relative
error $|\hat{w}/w-1|$ of the above approximation stays below $5\%$ and
it is asymptotically correct in the most important limit
$|\eps|\rightarrow 0$.

\section{Parameter region for relevance and reflected ion fraction}
\label{sec:apprefl}

\begin{figure}
\centering
\includegraphics[width=0.4\columnwidth]{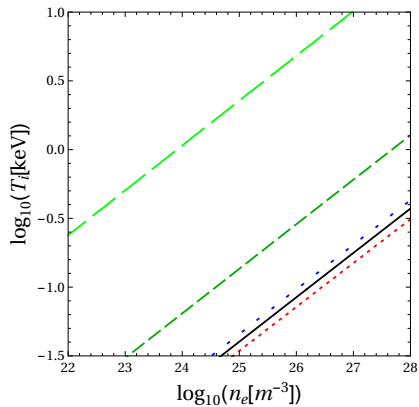}
\centering
    \caption{Approximate boundaries of the parameter regions where
      collisions qualitatively affect the dynamics of shocks. Below
      the lines $\nus > 0.01$, thus for shocks which live $100
      \lD/\cs$ the cumulative effect of collisions becomes order
      unity. Baseline parameters (corresponding to solid line): $Z=1$,
      $\tau=100$. Dashed: $Z=2$, long dashed: $Z=10$, dotted:
      $\tau=1000$, small dotted: $\tau=10$. }
    \label{fig:params}
\end{figure}

The importance of the fact that collisions act to populate ion trapped
regions is that the effect of collisions accumulates with
time. Therefore, even when $\nus\ll 1$ -- tempting one to neglect
collisions -- if the lifetime of the shock corresponds to an order-unity
collisional age, the shock will be significantly affected. As a rough
estimate for when collisions are relevant, in figure~\ref{fig:params}
we plot $\nus=0.01$ curves in the $n_e$--$T_{\ii}$ parameter space for
various values of $\tau$ and $Z$, which corresponds to $\nus t=1$ for
a reasonable shock lifetime of ${\sim}100\lD/\cs$. The $T_{\ee}$
dependences of $\lD$ and $\cs$ cancel, so only a very weak $T_{\ee}$
dependences remains (see the dotted curves) due to that of the Coulomb
logarithm. Also, the $m_{\ii}$ dependence of $\nu_{\ii\ii}$ and $\cs$
cancel, so it does not matter whether the charge-to-mass ratio is
hydrogen or deuterium like.  As expected, there is a rather strong
dependence on $Z$, which is relevant for non-hydrogen targets. For a
concrete example, in a hydrogen plasma with the parameters
$n_e=10^{27}\,\mathrm{m}^{-3}$, $T_{\ii}=100\,\mathrm{eV}$ and
$T_{\ee}=1\,\mathrm{MeV}$, a case with relevance to ion acceleration
that was considered in \citep{fiuza}, $\nus t$ becomes $1$ at time
$31\lD/\cs\sim0.7\,\mathrm{ns}$, that is shorter than the lifetime of
the shock.

While $\nus t\ll 1$, and so our model is valid, the shock properties
do not change qualitatively, so the effect of collisions on $\phimax$
is rather weak. The reflected ion fraction is
\begin{equation}
\alpha=\frac{\int_0^{v_0} f(v, \phi=0) \,\dd{v}}
{\int_\infty^{0} f(v,\phi=0) \,\dd{v}}
=\frac{\erf[\sqrt{\frac{\tau}{2}}\Ma]
  +\erf[\sqrt{\frac{\tau}{2}}( \sqrt{2\phimax}-\Ma)] }
{1+\erf[\sqrt{\frac{\tau}{2}}\Ma ]},
\end{equation}
where $f$ is the upstream distribution and $\erf$ denotes the error
function. For $\tau\gg 1$, this result reduces to
\begin{equation}
\alpha \approx \frac{1}{2}
\erfc\qty[\sqrt{\frac{\tau}{2}}\Ma\qty(1-\sqrt{F})],
\end{equation}
where $F=2\phimax/\Ma^2$ and it takes values close to $1$ for most cases
of interest, as discussed in \citep{pusztaishock}.

\begin{figure}
\centering
\includegraphics[width=0.4\columnwidth]{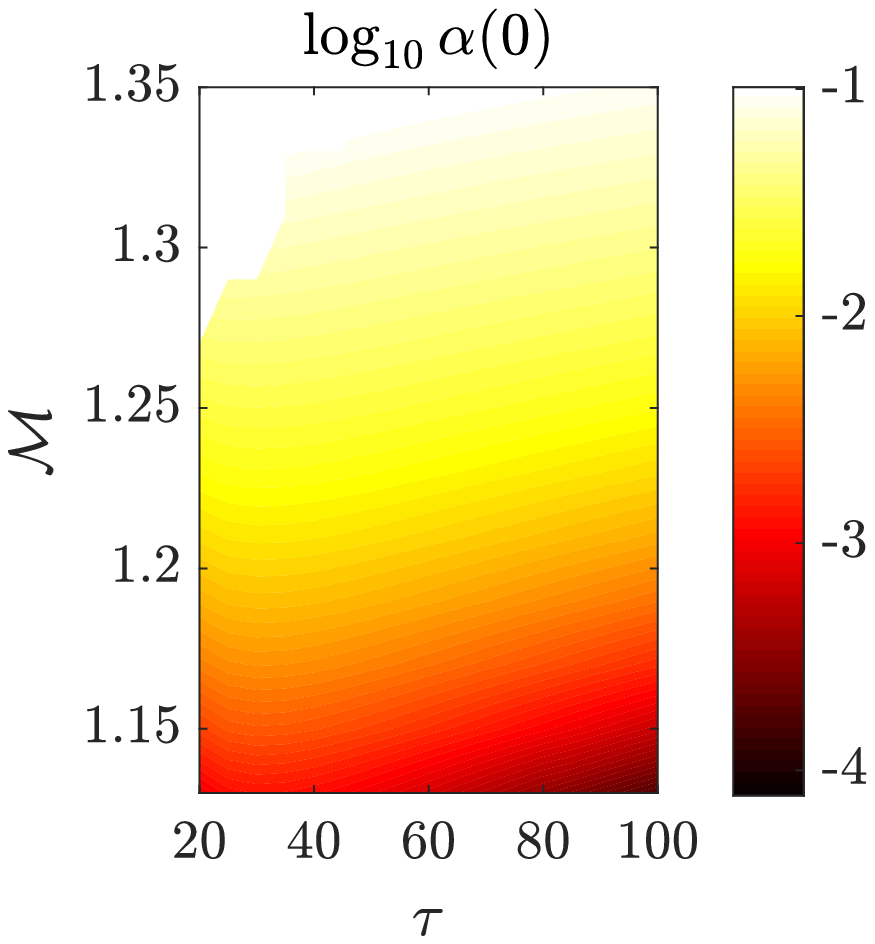}
\put(-115,130){\large (a)} 
\includegraphics[width=0.4\columnwidth]{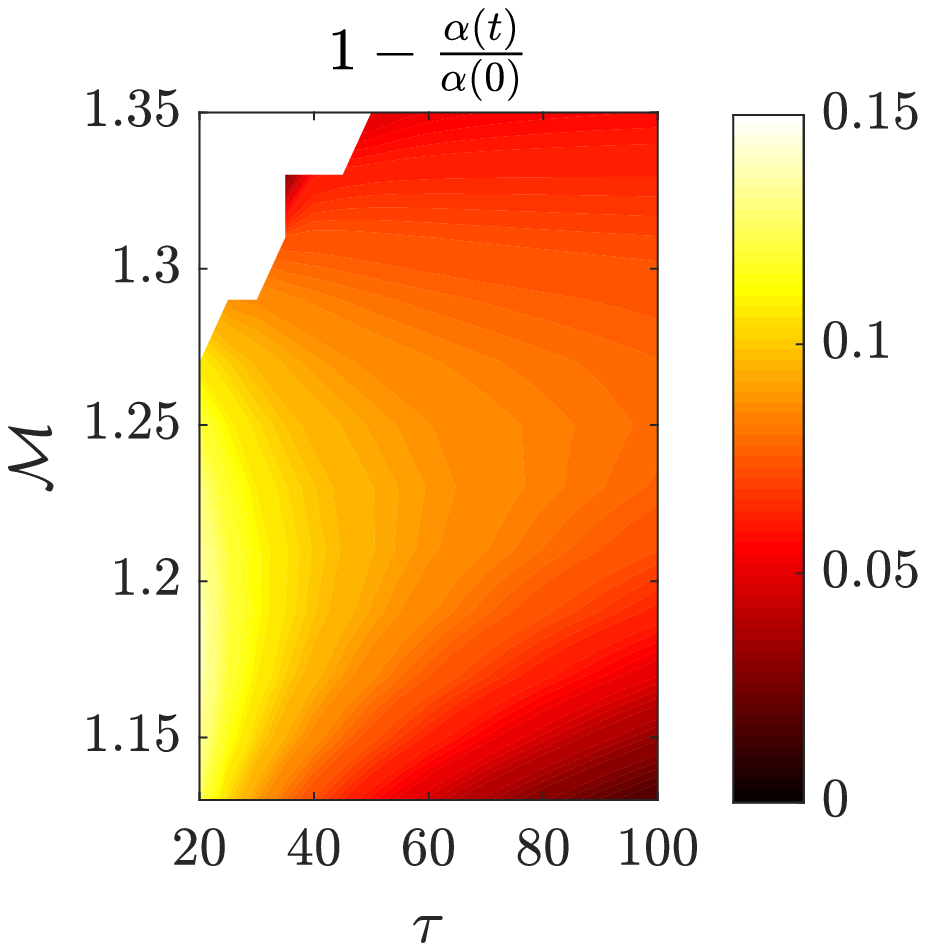}
\put(-115,130){\large (b)} 
\centering
    \caption{(a) Reflected ion fraction plotted as
      $\log_{10}\alpha(t=0)$, as a function of $\tau$ and $\Ma$. (b)
      Relative reduction in the reflected ion fraction for a
      collisional age of $\nus t=0.1$. }
    \label{fig:alpha}
\end{figure}

We find that $\alpha$ increases strongly with $\Ma$ -- as seen in
figure~\ref{fig:alpha}a -- since $\phimax$ increases stronger than
$\Ma^2$, corresponding to higher values of $F$. In turn, $\alpha$ is
exponentially sensitive to $F$. We have found that $\phimax$ is
reduced by collisions, and we might expect that the exponential
sensitivity to $\phimax$ can lead to significant changes in
$\alpha$. However, the reduction in $\phimax$ is so weak that even for
a collisional age of $\nus t=0.1$, the maximum relative reduction in
$\alpha$ is below $15\,\%$ in the region of parameter space shown in
\ref{fig:alpha}b. This effect is even weaker for higher $\tau$, which
is more relevant for ion acceleration experiments.

\bibliographystyle{jpp}
\bibliography{CollShocks}

\end{document}